\documentclass[AMA,STIX1COL]{WileyNJD-v2}
\usepackage{moreverb}

\newcommand\BibTeX{{\rmfamily B\kern-.05em \textsc{i\kern-.025em b}\kern-.08em
T\kern-.1667em\lower.7ex\hbox{E}\kern-.125emX}}

\articletype{Preprint}%

\received{Preprint}
\revised{Preprint}
\accepted{Preprint}

\usepackage{natbib}
\usepackage{lscape,rotating}
\usepackage{float,epsfig}
\usepackage{subfigure}
\usepackage{setspace}

\usepackage{amsmath} 
\usepackage{amsfonts}
\usepackage{amssymb}

\usepackage{lscape,rotating}
\usepackage{float,epsfig}
\usepackage{subfigure}
\usepackage{setspace}
\usepackage{amsmath}
 \usepackage{amsmath,bm} 
\usepackage{amsfonts}
\usepackage{amssymb}

\usepackage{amsmath}

\newcommand{\bc}{\begin{center}}
\newcommand{\ec}{\end{center}}
\newcommand{\be}{\begin{enumerate}}
\newcommand{\ee}{\end{enumerate}}
\newcommand{\bi}{\begin{itemize}}
\newcommand{\ei}{\end{itemize}}
\newcommand{\bt}[1]{\begin{tabular}{#1}}
\newcommand{\et}{\end{tabular}}
\newcommand{\bd}{\begin{description}}
\newcommand{\ed}{\end{description}}
\newcommand{\bal}{\begin{align}}
\newcommand{\eal}{\end{align}}
\newcommand{\bals}{\begin{align*}}
\newcommand{\eals}{\end{align*}}

\newcommand{\bes}{\begin{eqnarray*}}
\newcommand{\ees}{\end{eqnarray*}}
\newcommand{\bs}{\begin{slide}{}}
\newcommand{\es}{\end{slide}}

\newcommand{\mbf}[1]{\mbox{\boldmath$#1$}}

\newcommand{\bbmp}{\begin{boxedminipage}}
\newcommand{\ebmp}{\end{boxedminipage}}

\newcommand{\bY}{\mbox{\bfseries Y}}

\newcommand{\balph}{\mbox{$\mbf{\alpha}$}}

\newcommand{\bgam}{\mbox{$\mbf{\gamma}$}}

\newcommand{\bmu}{\mbox{$\mbf{\mu}$}}

\begin{document}

\title{Multiple imputation of partially observed data after treatment-withdrawal \protect}

\author[1]{Suzie Cro}

\author[2]{James H Roger}

\author[2,3]{James R Carpenter}


\address[1]{\orgdiv{Imperial Clinical Trials Unit}, \orgname{Imperial College London}, \orgaddress{\state{London}, \country{United Kingdom}}}

\address[2]{\orgdiv{Medical Statistics Department}, \orgname{London School of Hygiene \& Tropical Medicine}, \orgaddress{\state{London}, \country{United Kingdom}}}

\address[3]{\orgdiv{MRC Clinical Trials Unit @ UCL}, \orgname{UCL}, \orgaddress{\state{London}, \country{United Kingdom}}}

\corres{Suzie Cro, \email{s.cro@imperial.ac.uk}}

\presentaddress{Imperial Clinical Trials Unit, Stadium House, 68 Wood Lane, London, W12 7RH}

\abstract[Abstract]{
The ICH E9(R1) Addendum (International Council for Harmonization 2019) suggests
treatment-policy as one of several strategies for addressing 
intercurrent events such as treatment withdrawal when defining an estimand.
This strategy requires the monitoring of patients and collection of primary
outcome data following termination of randomized treatment. However, when patients
withdraw from a study before nominal completion this creates true missing
data complicating the analysis. One possible way forward uses multiple imputation to replace the
missing data based on a model for outcome on and off treatment prior to study
withdrawal, often referred to as retrieved dropout multiple imputation. This article explores a novel approach to parameterizing this imputation model so that those parameters which may
be difficult to estimate have mildly informative Bayesian priors applied during the imputation stage. A core reference-based model is combined with a compliance model, using both on- and off- treatment data to form an extended model for the purposes of imputation. This alleviates the problem
of specifying a complex set of analysis rules to accommodate situations where
parameters which influence the estimated value are not estimable or are poorly
estimated,  leading to unrealistically large standard errors in the
resulting analysis.
}



\maketitle

\section{Introduction}
\label{sec:Intro}
In pivotal clinical trials with longitudinal follow-up it is now common practice to attempt to collect full outcome
data after the withdrawal of randomized treatment up until the nominal end of study. This allows the study to readily adopt  a treatment-policy strategy for handling treatment withdrawal, to target the effect of the randomized treatment regardless of any treatment non-adherence \cite{iche9r119}. As discussed by Wang et al \cite{Wang23}, for regulatory decision making a treatment policy strategy is most often utilised in order to estimate how well a treatment will work in clinical practice. This follows  an intention-to-treat (ITT) ethic. Henceforth, we denote the intercurrent event of randomized treatment withdrawal  as
\emph{deviation} following Carpenter et al \cite{carpenter13}. If all post-deviation data are collected estimation of an estimand using a treatment policy strategy will be straight forward; standard analytical methods (e.g. ancova) can be applied to the observed data.
 However after deviation, collection of
 complete data is often problematic.
It is common for as many as half of those patients who deviate, to
subsequently withdraw from the trial and then have incomplete outcome data up to
and importantly at the end of their study period. Thus most often, estimation requires appropriate assumptions for the unobserved post-deviation data, which critically should align with the treatment policy strategy (i.e. reflect what actually happens to outcomes after treatment deviation). Fletcher et al \cite{fletcher22} on behalf of the EFSPI Estimand Implementation Working Group (EIWG)
suggest that ``there is little statistical literature concerning the unbiased
estimation of estimands using the treatment policy strategy, and this is an area requiring significantly more focus and attention.''

There are typically two main pathways for deviating trial patients; either progress to a pre-specified follow-on treatment (or no-treatment) and continue to be observed, or immediately leave the study with no further participation. Usually the follow-on
treatment will be the same for each arm, as otherwise this distinction becomes
an essential part of the treatment comparison. This shared follow-on treatment regime
may be the same or
similar to practice in the placebo or reference arm, suggesting that
those who discontinue in the placebo arm will experience little or no change in outcome.
But on the other hand the follow-on treatment may be 
less prescriptive than on-treatment reference including scope for rescue.
This aspect of the trial protocol is just as important when modelling
off-treatment outcomes as when interpreting the treatment-policy estimand.
When understood, it allows the generation of realistic models for unobserved outcomes during the follow-on period.

Three main ways forward have been proposed for handling partially observed post-deviation data in a longitudinal trial setting when adopting a treatment-policy strategy, mostly based on multiple imputation. First to ignore compliance status
and treat the data as a simple missing-at-random missing data problem with the intercurrent event
of treatment withdrawal ignored. However, this approach as we expand further on below is not fully aligned with a treatment-policy strategy. Second, to use
reference-based imputation (RBI) and assume the deviators behave like those observed in a specified reference group of the trial (typically placebo/control). This approach ignores any off-treatment data in the imputation and can then merge any
observed off-treatment data back in.  Third, model the
observed off-treatment and on-treatment data in a combined model and then impute
 the missing data based on this estimated model. Different variants of this
 latter approach, often referred to as retrieved drop out multiple imputation, use either imputation at final visit only, or a
 repeated measures based imputation across all visits, with the added complexity
 of differing choices for the repeated measures means model. These three
 ways forward and their variants are discussed in more detail in the following sections.

 Approaches based around RBI disregard potential information about
the outcomes off-treatment during imputation and also make strong assumptions
that may not always be consistent with trial experience in the active arm
after treatment withdrawal. Retrieved drop out approaches based on modeling outcome after treatment
withdrawal may be impractical because parameters are, or may be, poorly
estimated due to the limited observed off-treatment data post-deviation \cite{Wang23}.
Also it may be difficult to pre-specify an appropriate off-treatment model that
reflects the complexities whilst allowing the parameters to be
 reasonably well estimated.
 
In this article we propose a novel multiple imputation approach that draws its influences from
these two disparate approaches. The idea is to have a simple core model very
similar to those used in reference-based imputation. There the parameters for
the off-treatment period are stolen from those for the reference arm, usually
placebo. This core model is then extended with a series of additional parameters so that
the extended model is equivalent to the type of model used for modelling
off-treatment outcomes as previously described. 
The extended model is fitted using a Bayesian framework with typically uninformative
priors for the core model and mildly informative priors for the additional
parameters. In this way when there is little observed off-treatment data the subsequent
imputed data will behave somewhat similar to reference-based imputation but with
some slight increase in variation representing uncertainty about the RBI
assumption. Indeed some will see this as an improvement over RBI with its
total acceptance for the RBI's assumptions.  On the other hand,
when there is a large proportion of observed off-treatment data available this will over power 
the mild priors and the imputation will behave like the chosen off-treatment model.

This general concept can be applied to different kinds of outcome data including
binary, counts, time-to-event as well as general quantitative outcomes. For
simplicity of description, in this paper we focus on the case of a classic
repeated measures multivariate Normal model, with two treatment arms 
and possible baseline measurements to be treated as covariates.

In the next section \ref{sec:example} we describe two motivating trial data sets, which have identical data where observed as well as the same level of  off-treatment data post-deviation, but with missingness distributed differently between arms. The goal is to address an estimand using a treatment policy strategy to handle treatment deviation, but one data set is problematic while the other is not. We then review the current methods to handling partially observed treatment withdrawal data in the context of estimation for a treatment policy strategy in more detail in section \ref{sec:Methods}. Our novel approach, which builds upon the current methods, is described in full in section \ref{sec:Model1}. In section \ref{sec:ROTM} we apply our novel method and existing methods to the two example data sets and finish with a discussion in section \ref{sec:Discussion}.

 \section{Example data sets}
\label{sec:example}

To illustrate our approach we explore two publicly available example data sets based on real underlying  data from an anti-depression trial, originally conducted by Goldstein et al \cite{goldstein2004}, with the addition of simulated partial off-treatment data. The trial's primary outcome is change from baseline in Hamilton 17-item rating scale for depression (HAMD17) which was measured at baseline, 1, 2, 4, 6 and 8 weeks. Mallinckrodt et al (2013) extracted the on-treatment data from the original trial including a placebo group and an active group randomly chosen from three non-placebo arms: two doses of an experimental medication and also
 a medication approved at the time \cite{Mallinckrodt13}.  Data for the final visit, week 8, which was the original primary endpoint were not included and have never been made publicly available. For the purpose of this analysis the week 6 results have been treated as the trial's primary endpoint and also define trial completion . Completion rates (on-treatment) were 76\% (64/84) for active and 74\% (65/88) for placebo and the profile of visit-wise mean changes for patients
that completed the trial versus those who discontinued
early are summarized in their Figure 1.  
Michael O'Kelly and Sylvia Li extended this data set with simulated off-treatment data whilst exploring the properties of differing analysis methods and made available template SAS code through the Drug information Association Scientific Working Group on Estimands and Missing Data (SWGEMD)\cite{SWGEM} (row 3 in table \ref{LSPCM}). Deposited with the templates and freely available for download are the two data sets chosen to illustrate the methods at work and also to indicate the problems that occur when parameters in the model are non-estimable  \cite{SWGEM}.

For both data sets, the estimand of interest is the mean difference in the change in HAMD17 at week 6 (visit 4) among the eligible trial population with depression from active treatment versus placebo regardless of whether all treatment was received (treatment policy). Four patients had baseline less than 7 not indicative of depression, whilst 77 were in the mild range (7-17), 73 in the moderate range (18-24) and 18 severe (over 24). HAMD17 scores are expected to fall over the six weeks even in the placebo arm. The final column for placebo in Tables \ref{Missingness3} \ref{Missingness4} shows means for those who complete the trial on-treatment, with a reduction in mean of about one unit per visit. A larger drop of about 2 units per visit is seen for the completers in the active arm. On-treatment data are fully observed and both data sets have the same number of
on-treatment patients at each visit in each arm.
Off treatment data were simulated using a model estimated from the control group data only, conditional on a patient’s baseline attributes (baseline score and a pooled-site indicator) and also their observed on-treatment post-baseline outcomes. In terms of reference based imputation this is copy-reference (CR).The extent of missing off-treatment data is summarized here in table \ref{Missingness2} with all patients on treatment at the first visit. In these data sets all patients who deviate either stop follow-up immediately or continue to the end of the study period with complete observed off-treatment data.  
The first labelled \emph{converging} by the authors is called \emph{covered} here to indicate that it contains data which "cover" all subgroups defined by the models. This requires that it has off-treatment data observed at least once at each time point in all treatment by deviation-visit subgroups. As shall be later demonstrated this data set works well with all the multiple imputation approaches explored in this article. The second labelled \emph{nonconverging} by the authors and called \emph{perforated} here has no observations after deviation for any of the six patients in the active arm who deviate at week 2. This results in non-estimable parameters for some of the models discussed later. 
The \emph{perforated} data set has the missingness arranged in such a way as to generate a hole in the data. Since the on-treatment data in both 
data sets are based on the same original underlying data set, similar but not identical results are to be expected. Tables \ref{Missingness3} and \ref{Missingness4} summarize the means by visit for each combination of treatment, deviation visit and presence of off-treatment observation in the two data sets. In essence to estimate the treatment-policy estimand the blank cells in the visit 4 row need to be filled out and then the difference in means taken across columns for placebo, and then active, weighted by the count row.

\begin{table}
\caption{\label{Missingness2}Summary of on-treatment, observed off-treatment  and
unobserved off-treatment frequencies by visit in SWEGMD example data sets}
\begin{tabular}{|l|c|c|c|c|c|c||c|c|c|c|c|c|}\hline & \multicolumn{6}{c||}{\emph{\footnotesize{Covered}}} 
 & \multicolumn{6}{c|}{\emph{\footnotesize{Perforated}}}
 \\ \hline
 & \multicolumn{3}{c|}{\scriptsize{Placebo}}
 & \multicolumn{3}{c||}{\scriptsize{Active}}
  & \multicolumn{3}{c|}{\scriptsize{Placebo}}
 & \multicolumn{3}{c|}{\scriptsize{Active}}
 \\ \hline
  \footnotesize{Visit} & \scriptsize{On} & \scriptsize{Off-obs} & \scriptsize{Off-Miss} 
& \scriptsize{On} & \scriptsize{Off-obs} & \scriptsize{Off-Miss}
& \scriptsize{On} & \scriptsize{Off-obs} & \scriptsize{Off-Miss}
& \scriptsize{On} & \scriptsize{Off-obs} & \scriptsize{Off-Miss}
\\ \hline
 1 & 88 &  0 &  0 & 84 &  0 &  0 & 88 &  0 &  0 & 84 &  0 &  0  \\ \hline
 2 & 81 &  3 &  4 & 78 &  4 &  2 & 81 &  5 &  2 & 78 &  0 &  6  \\ \hline 
 3 & 76 &  7 &  5 & 73 &  5 &  6 & 76 &  9 &  3 & 73 &  4 &  7  \\ \hline 
 4 & 65 & 12 & 11 & 64 & 10 & 10 & 65 & 12 & 11 & 64 & 10 & 10  \\ \hline 
 \end{tabular}
\end{table}

\begin{table}
\caption{\label{Missingness3}Summary of on-treatment (non-bold) and observed off-treatment means (bold) by pattern of deviation and missingness in SWGEMD \emph{Covered} data set}
\begin{tabular}{|l|c|c|c|c|c|c|c||c|c|c|c|c|c|c|}\hline
& \multicolumn{7}{c||}{\emph{\scriptsize{Placebo}}} 
 & \multicolumn{7}{c|}{\emph{\scriptsize{Active}}}
 \\ \hline
& \multicolumn{7}{c||}{\scriptsize{Last Visit on treatment}} 
 & \multicolumn{7}{c|}{\scriptsize{Last Visit on treatment}}
 \\ 
& \multicolumn{2}{c|}{ 1} & \multicolumn{2}{c|}{ 2} & \multicolumn{2}{c|}{ 3} &  4
& \multicolumn{2}{c|}{ 1} & \multicolumn{2}{c|}{ 2}& \multicolumn{2}{c|}{ 3} & 4
 \\ \hline
  \scriptsize{Visit} & \scriptsize{Stop} & \scriptsize{Cont.} & \scriptsize{Stop} 
& \scriptsize{Cont.} & \scriptsize{Stop} & \scriptsize{Cont.}
& \scriptsize{Cont.}
& \scriptsize{Stop} & \scriptsize{Cont.} & \scriptsize{Stop}
& \scriptsize{Cont.} & \scriptsize{Stop} & \scriptsize{Cont.}
& \scriptsize{Cont.}
\\ \hline
 1 & 1.25 &  -1.67 &  -6.00 & -0.50 &  -1.00 &  -0.20 & -1.82 & 
 -2.00 &  1.25 &  -2.00 &  -10.00 &  -1.75 &  -3.40 & -1.75 \\ 
\cline{2-3} \cline{9-10} 2 &  &  \textbf{-2.29} &  -4.00 & 2.00 &  -2.83 &  -0.80 & -3.11 & 
  &  \textbf{0.19} &  -4.00 &  -4.00 &  0.00 &  -5.20 & -4.87 \\ 
\cline{4-5} \cline{11-12} 3 &  &  \textbf{-4.80} &   & \textbf{-0.97} &  -2.67 &  -0.80 & -4.45 & 
  &  \textbf{-2.71} &  &  \textbf{-5.00} &  -0.50 &  -6.00 & -7.25 \\ 
\cline{6-7} \cline{13-14} 4 &  &  \textbf{-4.21} &   & \textbf{-1.32} &   &  \textbf{-2.44} & -5.14 & 
  &  \textbf{-2.30} &  &  \textbf{-7.01} &   &  \textbf{-6.39} & -8.34 \\ \hline
\scriptsize{Count}  & 4 &  3 &  1 & 4 &  6 &  5 & 65 &  2 &  4 &  4 &  1 &  4 &  5 & 64 \\ \hline
 \end{tabular}
 \footnotesize{
Patients either stop follow-up at the deviation or continue being observed while off treatment until the end of trial. }
\end{table}

\begin{table}
\caption{\label{Missingness4}Summary of on-treatment (non-bold) and observed off-treatment means (bold) by pattern of deviation and missingness in SWEGMD \emph{Perforated} data set}
\begin{tabular}{|l|c|c|c|c|c|c|c||c|c|c|c|c|c|c|}\hline
& \multicolumn{7}{c||}{\emph{\scriptsize{Placebo}}} 
 & \multicolumn{7}{c|}{\emph{\scriptsize{Active}}}
 \\ \hline
& \multicolumn{7}{c||}{\scriptsize{Last Visit on treatment}} 
 & \multicolumn{7}{c|}{\scriptsize{Last Visit on treatment}}
 \\ 
& \multicolumn{2}{c|}{ 1} & \multicolumn{2}{c|}{ 2} & \multicolumn{2}{c|}{ 3} &  4
& \multicolumn{2}{c|}{ 1} & \multicolumn{2}{c|}{ 2}& \multicolumn{2}{c|}{ 3} & 4
 \\ \hline
  \scriptsize{Visit} & \scriptsize{Stop} & \scriptsize{Cont.} & \scriptsize{Stop} 
& \scriptsize{Cont.} & \scriptsize{Stop} & \scriptsize{Cont.}
& \scriptsize{Cont.}
& \scriptsize{Stop} & \scriptsize{Cont.} & \scriptsize{Stop}
& \scriptsize{Cont.} & \scriptsize{Stop} & \scriptsize{Cont.}
& \scriptsize{Cont.}
\\ \hline
 1 & -1.00 & 0.40  &  -6.00 & -0.50 &  0.13 &  -2.67 & -1.82 & 
 0.17 &   &  -5.00 &  -3.25 &  -0.67 &  -3.67 & -1.75 \\ 
\cline{2-3} \cline{9-10} 2 &  &  \textbf{-0.67} &  -4.00 & 2.00 &  -0.75 &  -5.00 & -3.11 & 
  &   &  -7.00 &  -3.25 &  -1.33 &  -3.67 & -4.87 \\ 
\cline{4-5} \cline{11-12} 3 &  &  \textbf{-2.93} &   & \textbf{-0.97} &  0.00 &  -6.67 & -4.45 & 
  &  &  &  \textbf{-4.18} &  0.33 &  -5.50 & -7.25 \\ 
\cline{6-7} \cline{13-14} 4 &  &  \textbf{-2.97} &   & \textbf{-1.32} &   &  \textbf{-7.25} & -5.14 & 
  &  &  &  \textbf{-5.29} &   &  \textbf{-6.20} & -8.34 \\ \hline
\scriptsize{Count}  & 2 &  5 &  1 & 4 &  8 &  3 & 65 &  6 &  0 &  1 &  4 &  3 &  6 & 64 \\ \hline
 \end{tabular}
  \footnotesize{
Patients either stop follow-up at the deviation or continue being observed while off treatment until the end of trial. }
\end{table}

 \section{Current methods}
\label{sec:Methods}
 
In this section we now review in more detail the three main approaches proposed to date for estimating an estimand with a treatment policy strategy for handling deviation.

\subsection{Ignoring compliance}
\label{sec:ignore}
The first approach assumes all outcomes are missing-at-random (MAR) ignoring whether a patient is on- or off- treatment. An argument for this approach of ignoring compliance centres on
the ITT ethic of ignoring everything except the randomized allocation and the final outcome.  It
focuses on hypothesis testing and suggests that all patients should be treated
as if termination of active treatment had never occurred and randomization is the
important criterion. But this does not fully align with  a treatment-policy
approach; post-deviation data are modelled on a combination of on- and off-treatment data from the patients randomized arm. Guizzaro (2020) \cite{Guizzaro21} used  causal inference methods to indicate
the importance of including an indicator for compliance when a treatment policy strategy is implemented, when the intercurrent
event of discontinuing treatment is not MAR. 

Despite its inherent failings this
approach has been implemented using a classic MMRM analysis either by maximum
likelihood (for example using the MIXED procedure in SAS) or Multiple imputation using Rubin's rules for inference (for example using the REGRESSION statement in the
MI procedure in SAS) \cite{Polverejan20} \cite{SWGEM}  \cite{EIWG} \cite{Roger17}. Since there is an incoherence in modelling off-treatment data based on a mixture of on- and off-treatment data for a treatment policy strategy it is generally unacceptable \cite{Roger19}. It should  only be used in situations where very few missing data are expected to avoid deviating from meaningful estimation. However there is no consensus on what constitutes very few.
 This use of on-treatment data to estimate off-treatment experience
will be especially problematic when off-treatment patient care allows wider treatment
options such as novel treatments.
The remaining approaches incorporate the compliance aspect in a series of different
ways.

\begin{table}
\caption{\label{LSPCM}Compliance based models for handling partially observed data after treatment withdrawal. For a continuous outcome $Y_{ij}$ measured for patient $i$ at times $j=1,...,J$, where $k$ is the last observation time prior to treatment discontinuation (i.e. deviation) and $\mu_{t,j}$ is the mean for treatment arm $t$ at time $j$}.
\begin{tabular}{|p{5.2cm}|p{5.2cm}|p{5.2cm}|}\hline
 Current\footnotemark[1] & Historic\footnotemark[2] & Full-pattern\footnotemark[3] \\ \hline
   $E[Y_{tj}]$= $\mu_{tj} \leq k $  \newline $E[Y_{tj}]=\alpha_{tj}$ for $ j>k$ \newline $\alpha_{tj}$ is the mean at visit j for the stratum defined by treatment
t and being off treatment at visit k &
$E[Y_{tj}]=\mu_{tj} \leq k $  \newline $E[Y_{tj}]= \gamma_{tjk}$ for $ j>k $ \newline $\gamma_{tkj}$ is the mean at visit j for the stratum defined by treatment
t and the last-on-treatment visit k (history) & 
$E[Y_{tj}]=\gamma_{tjk} $ for all $j,k$ \newline  \newline $\gamma_{tkj}$ is the mean at visit j for the stratum defined by treatment
t and the last-on-treatment visit k (history)\\ \hline 
 \end{tabular}
 \\ \footnotesize{
1: Referred to as A3 by SWGEMD\cite{SWGEM}, M12 by EIWG  \cite{EIWG} , M2 by Roger (2017) \cite{Roger17} and rd\_trt and  rd\_trt\_dcreason by Polverejan \& Dragalin (2019) \\
2: Referred to as A2a by SWGEMD\cite{SWGEM} and M3 by Roger (2017) \cite{Roger17}   \\
3: Referred to as A2 by SWGEMD\cite{SWGEM}, M13 by EIWG  \cite{EIWG} and M4 by Roger (2017) \cite{Roger17} }
\end{table}

\subsection{Reference-based multiple imputation}
The second broad approach imputes all post-withdrawal data using a
reference-based multiple imputation (RBI) method such as jump-to-reference (J2R) ignoring
off-treatment data and then, where available, replaces imputed values with
actual observed measurements \cite{carpenter13}. The underlying model for reference-based multiple imputation of a  continuous outcome is the standard repeated measures multivariate normal model including the treatment by time interaction, along with other covariates, which may or may not be crossed with time, and an unstructured covariance matrix that may be shared or separate for each arm. For imputation, a model is constructed for each pattern of withdrawal using the parameters based upon the underlying fitted multivariate normal model. How the construction is performed defines the different reference based methods. Multiple imputations are then drawn following the Bayesian paradigm. As in standard MAR MI, post imputation each imputed data set is analysed with the substantive analysis model of interest and results combined with Rubin's rules.  

Jump-to-reference imputation forms the imputation model using the same mean profile up to deviation as observed for the randomised arm, but post deviation the profile jumps to the estimated profile for the reference arm.  Other reference-based approaches include copy increment from reference (CIR) which forms the imputation model using the same mean profile up to deviation as observed for the randomised arm, but post deviation the profile tracks parallel to the pattern for a specified reference arm. It is also possible to implement different reference based assumptions for different individuals in the same trial based on reason for withdrawal. Polverajin \& Draglin \cite{Polverejan20} use copy increment
from reference (CIR) reflecting their interest in a long term degenerative
disease where the effect of treatment will remain and there is no bounce back to 
untreated values as in jump-to-reference. This they label CIR in contrast
to an analysis where the reference-based assumption varied by type of deviation (labelled BY\_REASON) where copy reference (CR) is used for AEs, J2R for loss of
efficacy (LOE) and CIR is used for the ``other'' category. 
Both the EIWG \cite{EIWG} and Wang \& Hu \cite{Wang22} use J2R as the RBI method when evaluating methods to target a treatment policy estimand, and there are now examples of this method being used in practice \cite{Tan21}.

A variant of this method, referred to as the ``return to baseline approach'' \cite{Wang23}, assumes there is no placebo effect and any intervention effect will be washed out after deviation.  Data are multiply imputed based on the mean baseline values \cite{Leurent20}. This approach washes out any drug effect after treatment deviation whilst assuming any post-deviation treatment prevents further deterioration from baseline. This can be viewed as a particular type of reference based multiple imputation where the reference is now the baseline mean. This contrasts with the baseline observation carried forward (BOCF) approach  where the baseline observation is simply carried forward; this latter approach is not recommended since it underestimates within patient variability of the measurements at different time points \cite{Qu22}. Others have suggested imputing based on the patients baseline observation plus some random variation \cite{Wang23} but this has shown to result in biased mean when the missingness depends on observed baseline and/or post-baseline intermediate outcomes and the variance of imputed values can be much larger than the variance of the baseline values \cite{Qu22}.

A  more simplistic reference-based approach termed the ``washout method'' by Wang et al \cite{Wang23} excludes intermediate measurements in the imputation model. An imputation model is built based on placebo/reference completers only and  the missing observations for those in the active arm at the primary endpoint are directly imputed without imputing any intermediate values. For patients in the placebo/reference arm the MAR assumption is accepted and intermediate outcomes are included in the imputation model. As described by Wang et al \cite{Wang23} this approach washes out the experimental treatment effect in those who drop out after treatment deviation in the experimental/active treatment group. 
  As Wang et al \cite{Wang23} suggest as an aside in the appendix "the washout approach can be thought as a simplified version of the J2R approach without considering the intermediate measurements from any arms in the imputation".

\subsection{Retrieved drop out multiple imputation}
\subsubsection{Compliance model - last visit only}
Wang \& Hu \cite{Wang22} propose a novel approach using multiple imputation 
of the outcome variable at final visit imputed  for those who terminate prior to
 final visit without imputing earlier missing data. This imputation uses a regression
for final outcome on patient's baseline and their last on-treatment value,
stratified by randomized treatment and the visit number of last on-treatment visit.
Intermediate outcome measurements either before or after the last on-treatment
visit are ignored, as well as ignoring those who complete on treatment.

An alternative approach, (the fourth approach  in the SWEGMD templates \cite{SWGEM}, labelled A4) also imputes
missing values at the final visit ignoring those who complete the trial on
treatment. It assumes that the change in outcome from time of discontinuation of
treatment to end of scheduled follow-up is MAR in patients who discontinue study
treatment, conditioning only on baseline, last on-treatment outcome, and the
length of time on study treatment. The time on study is considered continuous
and the relationship between time on study and outcome is considered linear.
This is then similar to Wang \& Hu \cite{Wang22} but rather than
stratify by treatment and off-treatment visit number they regress on the numeric
 value of the visit. This reduces the number of required parameters in the 
regression. 

Focusing on imputation at final visit alleviates the need for imputing
intermediate missing values, but requires regressing on data that is
certainly observed, such as the outcome at or just before deviation
irrespective of when during the trial deviation occurred.
This approach obviously suits those studies where after treatment
withdrawal outcome is only measured at the final nominal visit.
  Also the details of the imputation process are likely to be less contentious. But the strong assumption of linearity between outcome and time on study may not always be appropriate. Further, when intermediate off-treatment data are extensive, such as in long term diabetes studies,
    there will presumably be some loss of information in discarding the intermediate data.

\subsubsection{Compliance model - repeated measures}
Here the outcome is modelled across all visits accounting for compliance using
all the observed data and some form of a repeated measures model.
A basis for such an approach is that prediction after deviation should rely only on observed data from those who deviate. However this can be too narrow an approach when available off-treatment data are sparse.
The simplest approach is to fit a MMRM model as
under MAR (see Section \ref{sec:ignore})  and add a single indicator variable that denotes whether the patient
discontinues treatment before the end of trial or not. Polverajin \&
Draglin \cite{Polverejan20} denote this MAR\_DC and also do the same ignoring the patients who
complete denoted RD\_SUBSET. The
indicator variable On-Off(Patient) is defined at the patient level and denotes discontinuation of
treatment at any time before the final visit.
As a result it can be fitted sequentially using software such as proc MI in SAS
and does not require regression on previous residuals. 

At the other extreme, the most complex repeated measures model effectively
stratifies by both treatment arm and the index of the last visit on treatment
and imputes within each stratum combination \cite{Wang23}.
The second approach in the SWEGMD templates (labelled A2) does something very similar
but whilst stratifying  by treatment it does
not cross baseline with treatment. Instead it uses a sequential
regression with baseline and an interaction between treatment and index of last
visit on treatment. In the purest form of complete stratification by arm, the study completers who are still on-treatment
are irrelevant in the imputation process as they form their own stratum with no
missing observations; here their data does not inform the imputation of the unobserved off-treatment data. This pure approach is MI3 from
EIWG \cite{EIWG} and M4 in Roger (2017) \cite{Roger17} (see table \ref{LSPCM}). As implemented by SWEGMD (A2) the completers have impact
on estimated baseline regression coefficients.  All the implementations effectively use a single 
covariance matrix with no attempt to fit separate covariance matrices either by arm or,
by before-deviation versus after-deviation as in Carpenter et al\cite{carpenter13}.
Both these models have the property that means differ depending upon whether the patient deviates or not at later visits.

If there are $J$ visits in the later A2 model, then potentially $J^2$ parameters are
introduced into the model, J for each regression, although in the purest form
many of these have no influence in the imputation process.
For large J this can be problematic.
An adapted version of their second approach, A2a, is proposed within the
SWGEMD code \cite{SWGEM} where each regression only depends on ``last visit'' number up to
the current one with all those reaching this far on treatment pooled in one level (see table \ref{LSPCM}).
 This is equivalent to the M3 method in Roger (2017) \cite{Roger17} . Effectively the strata or factor levels are defined by pooling all those
  patients who are on treatment at this visit. Indeed if
 baseline were absent from the model crossed with treatment
  then this pooling would have no impact on the subsequent imputation.
Whether pooled or not the series
of factors are nested within each other from regression to
regression, so again regressing on observed or residual is identical. This A2a approach we will call the \emph{historic} compliance model as it accounts fully for the compliance history so far as a whole, in contrast to the A2 approach described previously  which we refer to as \emph{full-pattern}.

All the same, the
final regression will potentially have J strata and some of these may be very small with little
outcome data observed at final visit. As long as an MAR assumption for study withdrawal
holds this approach will behave well for large data sets. Whenever the strata size
are small (i.e. few individuals with the specific deviation pattern) or study completion rate is low, sharing information between strata 
should improve precision at the potential expense of introducing bias. 

A simpler approach,  uses
an on/off treatment indicator at each visit within each arm but unlike the first
approach this indicator changes value from visit to visit and hence from
regression to regression. This means that the model at each visit is not nested
within the model at subsequent visits as the variable for on/off treatment changes from visit to visit. This is the third approach labelled A3 in the SWGEMD templates \cite{SWGEM} (see table \ref{LSPCM}). 
 It is similar to proposed methods
RD\_TRT and RD\_TRT\_DCREASON discussed in Polverajin \& Draglin \cite{Polverejan20}, MI2 form EIWG \cite{EIWG} and M2
in Roger (2017) \cite{Roger17}.
 In order to fit this model as a sequence of regressions each
regression in the sequence needs to regress on previous residuals rather than
on previous observed outcome values. In the SWEGMD SAS code this is
done using an intermediate call to the MIXED procedure between calls to the MI
procedure, or using the MISTEP macro from the DIA missing data web site \cite{SWGEM}. 
The simplifying assumption in this model is that the marginal impact of being
off treatment at the current visit is the same whether or not you were off treatment at
previous visits. We refer to this A3 approach as the \emph{current} compliance model as it accounts for whether or not the patient is currently on treatment.

A relatively large reduction in number of parameters can also be achieved by
regressing on the quantitative length of time on treatment rather than handling
deviation visit as a multilevel factor. This is implemented in the SMWEG A4
template code, which is discussed in the above section \cite{SWGEM}.

\section{Novel model parameterization}
\label{sec:Model1}

We now describe our novel approach which combines a core reference-based model with a compliance model to form an extended model for the purposes of imputation. 
Assume a total of N patients are randomized to two treatment
groups, reference and active. They are observed at baseline and J subsequent
visits, with interest in the outcome at the final visit J. Let $\bY_i = (Y_{i0},
Y_{i1}, \ldots ,Y_{iJ})$ denote the longitudinal vector of continuous outcome
values for the i'th patient and $T_i$ denote randomized arm taking values 0
for reference and 1 for active.
For simplicity, assume there are no interim missing data and all patients are observed at baseline
 and up to visit $D_i$, the last observation time prior to treatment
 discontinuation for patient i, known as deviation here. This is set to final
 visit $J$ for those completing on treatment. Patients who withdraw from treatment and also leave before the end of the study having 
 no final visit outcome measure, have their final visit observed within the
 study at visit $S_i$, which is set to $J$ for those completing study.
 Patients halt study treatment when leaving the study and so $S_i \geq D_i $ for
 all patients. Outcome values off treatment are assumed to be available between
 visits $(D_i+1)$ and $S_i$ inclusive for the i'th patient but missing after
 $S_i$. This set will be null when deviation and study withdrawal concur
 ($D_i=S_i$).
 The on-treatment model conditional on remaining on treatment has mean outcome 
 $E[Y_{ij} | D_i \geq j, T_i=t ] = \mu_{tj}$.
 
 Two RBI methods are commonly used and offer possible core models. First J2R is
 defined as
\begin{eqnarray}
\label{RBI}
E[Y_{ij} | D_i = k, T_i = t] &=& \mu_{tj} \mbox{ for } j\leq k \nonumber\\
                                &=& \mu_{0j}  \mbox{ for } j > k
\end{eqnarray}
whilst CIR is defined as
\begin{eqnarray}\label{CIR}
E[Y_{ij} | D_i = k, T_i=t] &=& \mu_{tj}  \mbox{ for } j\leq k\nonumber\\
                                &=& (\mu_{0j} +  \mu_{tk} -\mu_{0k})  \mbox{
                                for } j > k
\end{eqnarray}

J2R is suited to a treatment that is expected to
have only short term effect and where the response would return to that of the
reference after treatment withdrawal. On the other hand, CIR is more appropriate
for a treatment where the improvement obtained from treatment up to time of deviation will
continue, but further improvement relative to reference is unlikely. An
intermediate profile \cite{White20} or other shape based on the reference arm experience might be chosen for the core
means model. Three other potential core models are last mean carried forward (LMCF), where the mean stays the same as the last on-treatment mean for a patient, and return to baseline (RTB) where the off-treatment mean is the baseline mean within arm or across both arms, or missing at random (MAR) where the means are simply $E[Y_{ij} | T_i=t] = \mu_{tj}$. The core model could also include a fixed "delta" depending on visit  or number of visits since deviation as demonstrated in one of the later examples below. It is important that the choice of core model is based on clinical experience and plausibilty, rather than mathematical simplicity.

Note how the on-treatment parameters $\mu_{tj}$ can be estimated from on-treatment data and the  means for the off-treatment period via the core model (RBI or other) conditional upon the history (on/off treatment pattern so far). When data are available for the off-treatment period it is possible to fit a more complex model for this period (for $j>k$) by also using one of the suggested compliance models. This we call the \emph{extended} model. The idea is to centre the extended model on the core model by simply subtracting off the projected means from the core model. Then if the core model is a good description, the extra deviation part (difference between compliance model and core model) will be small in value.  When we expect holes in the data one way forward to make the compliance model, hence deviation parameter, estimable is to use a mildly informative zero-centred prior for the extension parameters. However the core model, which conditions upon the history (on/off treatment pattern so far), needs to be nested within the compliance model otherwise it will distort the fitted compliance model. That is when sufficient off-treatment data are available and the prior is made totally uninformative the compliance model ought to be recovered as a special case. Examples where nesting does not hold are discussed in the following section.

There is some choice for the compliance part of the extended model with the saturated compliance
model adding a treatment by history by visit interaction in the
off-treatment period.
\begin{eqnarray}
E[Y_{ij} | D_i = k, T_i=t] &=& \mu_{tj}  \mbox{ for } j\leq
k \nonumber\\
                                &=& \gamma_{tkj}  \mbox{ for }
                                j > k
\end{eqnarray}
where $\gamma_{tkj}$ is the mean at visit j for the stratum defined by treatment
t and the last-on-treatment visit k (history). This is the \emph{historic} compliance model, History*Treatment*Visit ,and is preferred in this context to the \emph{full-pattern} compliance model, Pattern*Treatment*Visit, where pattern is declared at the subject level and reflects the compliance throughout the trial. In contrast the history only reflects the compliance so far.

Formally, the proposed approach is to re-parameterize this extended model by including an additional offset term determined by a reference-based model for the post deviation means. No additional parameters are needed as the reference based mean involves parameters $\bmu$ from the on-treatment observations, and these parameters concur with those for the extended model. What changes is the interpretation of the parameters in the extended model. These are now deviations from the reference-based mean rather from a global mean of zero. For instance when J2R is used as reference-based core the extended model becomes
\begin{eqnarray} \label{gammastar}
E[Y_{ij} | D_i = k, T_i=t] &=& \mu_{tj}  \mbox{ for } j\leq
k \nonumber\\
                                &=& \mu_{0j} +  \gamma^*_{tkj}  \mbox{ for }
                                j > k.
\end{eqnarray}
Then a mildly informative
zero-centred prior is applied to the parameters $\bgam^*$ whilst retaining
an uninformative prior for $\bmu$ (the parameters from the core model).
Note how in equation \ref{gammastar} the same parameters $\bmu$ are shared between the on and
off-treatment periods by the J2R RBI model.
There is no aliasing of parameters between $\bmu$ and $\bgam^*$ 
as $\mu_{0j}$ for $j>k$ is inherently estimated by the data where $j<k$ in other patients.
When data are not available for the off-treatment period
then in expectation this reduces to the J2R reference based imputation as
the $\bgam^*$ priors have zero mean. This complex extended model would suit any other
core model such as CIR equally well. When off-treatment data are available then
$\bgam^*$ becomes estimable and in the Bayesian model some very small amount of information will be
fed back into $\bmu$ as $\bgam^*$ has a mildly informative prior.
 However this is not an issue as the model is only being used to impute missing
  data in the off-treatment region.
It is not being used to estimate treatment effect. Intermediate
missing values in the reference arm on-treatment data could still be handled as
missing here and used for imputation.

Less flexible options for the extended model could be chosen based on experience
with modelling retrieved dropout. An alternative simple choice is to
replace the parameters $\gamma_{tkj}$ by $\alpha_{tj}$ and $\gamma^*_{tkj}$ by $\alpha^*_{tj}$ in the same way as the \emph{current} compliance model based on On-Off(Visit)*Treatment*Visit where On-Off(Visit) is a visit level indicator for whether the patient is on or off treatment.
\begin{eqnarray}
E[Y_{ij} | D_i = k, T_i=t] &=& \mu_{tj}  \mbox{ for } j\leq
k \nonumber\\
                                &=& \alpha_{tj} = \mu_{0j} +  \alpha^*_{tj}  \mbox{ for }
                                j > k
\end{eqnarray}
Here $\alpha^*_{tj}$ is the mean at visit j for the stratum defined by treatment
t and the impact of being off-treatment on the expected outcome is the same however long the patient has been off-treatment. 
 Replacing the
interaction in the full-history model by two main effects is another option.
Linearization of $\bgam$ or $\bgam^*$, rather than using a distinct mean for each visit (in a similar way to A4 in table
\ref{LSPCM}), is a further possible simplification but neither is explored here.

\section{Robust off-treatment modelling: application to example data sets}
\label{sec:ROTM}
When there is sufficient data within patient strata defined by treatment and deviation visit (pattern) then it is relatively easy to define a compliance based multiple imputation procedure within strata that yields a robust estimator of treatment comparison at final outcome. Of course assumptions about the missingness process
conditional upon pattern will remain, a topic that is left until the discussion section. The \emph{historic} and \emph{current} choices for the compliance model offer
 less robust approaches which can be used with less extensive off-treatment data. 
 
 The
 \emph{perforated} data set from SWEGMD demonstrates the problems that can
 occur when exclusively using a compliance based modelling approach. Here with only four visits, seven patients on placebo and six on active deviate
 between first and second visits. Whilst five out the seven are observed on placebo,
 none of the six on active are observed (table \ref{Missingness2}). This means that 
 the parameters $\gamma_{112}$,$\gamma_{113}$ and  $\gamma_{114}$ for the \emph{historic} approach and
 $\alpha_{12}$ for the \emph{current} approach are not estimable. However these parameters are in general required for imputation of unobserved values. 
 Rather than supply informative priors for these parameters directly, we proposed above that an extended model should be re-expressed, as an appropriate RBI core model with extra deviation parameters, which are the difference between the required compliance model and the selected core RBI model. 
 Then mildly informative priors for these deviation parameters are applied which centre
 on zero. Here we focus on \emph{historic} and \emph{current} as  suitable choices for the compliance model used in the extended model.
 The later requires an assumption that having come off treatment the expected
 outcome is the same however long ago the deviation occurred. For the RBI core model 
 we use J2R and CIR as realistic options.

For the antidepressant trial example data we may expect those on placebo who stop treatment at a specific visit and subsequent follow-up to possibly
\begin{itemize}
    \item fair worse than the placebo mean as they stopped because they are severely depressed, or
    \item fair better than the placebo mean as they have improved and see no purpose in remaining in the study, or
    \item fair about the same, as the off treatment policy in the trial is the same as the placebo on-treatment policy.
\end{itemize}
 For the first two the effect may be controlled by regressing on baseline HAMD17  and previous outcome. That is by conditioning on previous depression state the future residuals may be simply MAR. A similar picture may hold for the active arm with the additional consideration of the withdrawal of a potentially active therapy. So some intermediate position between J2R and CIR may be a reasonable choice for the core model. As an example of using a fixed delta as core we show results for an MAR+2 core where the placebo arm is MAR whilst the active arm is MAR with 2 units added for each extra visit following deviation, representing a rather pessimistic core.

Using the Wilkinson-Rogers model formulae notation \cite{Wilkinson73},
there are two main terms in the model, one for the core model
and one for the deviations. Assume the following variables defined at the
patient-visit level; factor Trt is treatment as randomized (two levels here),
quantitative indicator OffT with 0 for on-treatment and 1 for off-treatment,
factor Pattern has value which is the current visit number whilst on treatment and the index of last
on-treatment visit (deviation) when not, factor J2R has active treatment level
whilst on treatment and otherwise reference treatment level. For more complex cores such as CIR (or LMCF) it is necessary to declare an array of treatment-by-visit  quantitative
variables TrtbyVis[2,4] reflecting for instance equation \ref{CIR} and then regress on these (see appendix for more detail). Neither of CIR, or (MAR+$\delta$) is nested with the \emph{current} compliance model and both require use of the fuller \emph{historic} extended model, as indicated by blank entries in table \ref{sec:results}.

The core part of the model is merely J2R*Visit for J2R or TrtbyVis1-TrtbyVis8 for  CIR.
The deviations are simply OffT*Trt*Visit*Pattern for \emph{historic} or simply
OffT*Trt*Visit for \emph{current}. Then any baseline covariates
such as Baseline can be added as required, such as in this complete model
\begin{eqnarray}
\label{J2R_A2a}
\mbox{J2R*Visit + OffT*Trt*Visit*Pattern + Baseline*Visit*Trt}. \nonumber
\end{eqnarray}
Note here how both second and third terms are regressions on a quantitative
variable, OffT or Baseline. Importantly when a subject is still on treatment
OffT is zero and the model reduces to J2R*Visit +
Baseline*Visit*Trt. This requires that OffT is not treated as a factor (not on the CLASS statement in SAS).

Using the SAS procedure proc BGLIMM this Bayesian model can be implemented directly allowing for
missing data. The repeated measures multivariate Normal is specified through the
REPEATED statement usually using an unstructured covariance matrix which could be grouped by treatment arm
if required. As long as conjugate priors are used, as here, the procedure uses
direct sampling resulting in very little serial correlation in the sampled
parameters.
However to take advantage of the proposal, mildly informative priors need to be
attached to the $\bgam^*$ or $\balph^*$ parameters. This is quite messy as the names
automatically allocated to the deviation parameters by the SAS procedure are long and include the
variable name and associated value for each element of associated term, for
instance ``OffT*Trt 1*visit 2*Pattern 1''. The example code attached 
to this paper demonstrates how this can be accomplished most easily.
Numeric factor values make this easier, but care must be taken as SAS names can be no longer than 32 characters.

The variance chosen for the prior for the deviation parameters will alter results,
varying between the RBI core and the extended model as extremes, and therefore needs to be carefully considered. We propose using the
residual variance of the outcome at final visit from a simple MMRM model of
on-treatment data.
The prior can then be seen as providing one extra patient-visit worth of additional likelihood in to the posterior for each parameter. This is one patient's worth for \emph{current} and slightly more for \emph{historic} depending on the number of visits.
For the example data sets used here we chose the variance to be 40 on this basis. This value could be chosen
during trial design based on previous experience and written into the protocol or analysis plan. Here
we show results for these two data sets with the variance for prior changing through
1, 10, 40, 160 and 1000 for each of a range of choices for model alongside results using classic J2R RBI using proc BGLIMM (with or without observed off-treatment data added)
and the standard \emph{historic} and \emph{current} compliance models without re-parameterization 
using three differing computational methods (proc MI, MISTEP macro or proc BGLIMM).  For the \emph{perforated} data set the latter three
computational routes give different answers based on how they cope with the non-estimable paramaters. For the SWGEMD proc MI route 
any non-estimable parameter value is set to zero in the imputation
stage. In the MISTEP macro it effectively removes any affected patient-visit combination by
setting the imputed value to missing. For the BGLIMM route using MCMC and
Bayesian parameters to represent missing values, we set very mild zero-mean
Normal priors for all the fixed effect parameters with variance set to one thousand
regardless of their role in the model.  We used 10,000 imputations which
leads to an approximate Monte Carlo SE of 0.002 for treatment difference, whilst our results are
presented to 2 decimal places. For all multiple imputation methods, results were combined across imputed data sets for inference using Rubin's rules.

\subsection{Results}
\label{sec:results}

The results are summarized in table \ref{Results};
 The left side uses the
\emph{covered} data set where all parameters in the \emph{historic} and \emph{current} models can be
estimated, whilst the right uses the \emph{perforated} data set
where difficulties in model fitting occur. The top two rows  show the simplest
approach using the RBI method J2R using on-treatment data, either on its own or
more sensibly with observed off-treatment data added back in.  

When J2R MI is implemented for the \emph{covered} data set, adding the
observed off-treatment data back to replace the imputed J2R data when available changes the estimated
mean difference from 2.18 to 2.28 and reduces
the standard error from 1.13 to 1.05 as expected.
This reflects how the ongoing treatment difference is partially maintained after deviation in the observed which contrasts with J2R where it is assumed the treatment effect immediately disappears following deviation.

For the \emph{perforated} data set the pattern in SE is similar but the mean difference increases by a larger factor from 2.17 with J2R MI only to 2.39 with J2R MI and observed off treatment data . This is because the placebo arm has more off-treatment observed data and less observed on-treatment data in the active arm (table \ref{Missingness2}), which increases the overall treatment-policy difference compared to the \emph{covered} data set. 

For the compliance based MI analysis, as we
expect, when all parameters in the \emph{historic} and \emph{current} compliance  models are estimable the
differing computational routes deliver the same result.
For the compliance based MI analysis using the \emph{covered} data set the simpler \emph{current} model
gives a smaller SE of 1.07 (or 1.06) compared to the fuller \emph{historic} at 1.10 with a small change in
mean from 2.34 (or 2.35) to 2.32 (or 2.31) when all parameters are estimated. For the
\emph{perforated} data set all three computational methods have to take
evasive steps for both the simpler \emph{current} and the \emph{historic} model. The stepwise approaches (MISTEP and Proc MI) fair worst as they need to impute at the intermediate visit 2 where the non-estimable parameter is required. MISTEP replaces any patient-visit combinations which
cannot be imputed by missing values and the mean becomes biased whilst the SE gets smaller
showing how off-treatment patients increase variability in the outcome variable for
the merged treatment-policy data set. That is those on and those off treatment at final visit 
are inherently different and reducing the number in the smaller subgroup decreases the variance in the 
merged set. The Proc MI procedure effectively sets the non-estimable $\alpha_{12}$ to zero and carries on. This is safe as it allows the imputation to proceed to subsequent visits with a full set of patients. But it makes an inherent assumption whilst most likely underestimating the SE.
The same picture applies to the BGLIMM computational approach where
all fixed effect parameters are given a broad prior with variance of 1000, with
the SE increasing in the \emph{historic} case where the non-estimable parameter $\gamma_{124}$
is required directly for imputation at the final visit.
Diagnosis of worm plots from the non-estimable parameters indicates roaming during MCMC. For the simpler \emph{current} model the BGLIMM results are valid and match the later results using the recommended route.

The extended core model using J2R as core with differing variances for priors behaves well with both data
sets. In the extreme with unit variance it behaves like J2R RBI with added
on-treatment data, whilst with larger prior variance for the deviations it
behaves like the associated off-treatment model \emph{historic} or \emph{current}. With the \emph{perforated} data set the SE goes up as the variance
goes up for J2R+\emph{historic}, but remains stable for J2R+\emph{current}. With large variance for prior J2R+\emph{historic} becomes unstable just like the \emph{historic} compliance model.

The CIR core model is only nested within the full \emph{historic} compliance model and should not be used with the \emph{current} compliance model. This is because the impact of having deviated under CIR depends upon the visit at which the patient deviated. Only results using the \emph{historic} approach are presented.
Using CIR rather than J2R as core model is expected to increase the estimated treatment difference as the treatment effect from the deviation visit is carried forward.
With the \emph{covered} data set this is evident when a stronger prior is used such as unit variance. With our recommended mild prior variance of 40, the increase with CIR is 2.36 compared to 2.32 for J2R. For the \emph{perforated} data set it should be remembered that large variances for the priors will lead to problems with the MCMC chain and so only variances 1 to 160 are of interest.
With prior variance of 1000 the SE is large whilst the non-estimable parameter roams in the MCMC chain. Using the recommended variance for prior of 40 the results for CIR+\emph{historic} model are very similar to those for J2R+\emph{historic}. 

When the core model is MAR with a delta of +2 for each subsequent visit after deviation, the mean treatment difference is 2.38, which is larger than for either J2R or CIR. This reflects a strong deleterious impact of withdrawal of active treatment in the core model.

 \begin{table}
\caption{\label{Results}Mean treatment policy difference and SE using MI for two
EIWSG similar data sets using differing methods.}
\begin{tabular}{|l|c|c|c|c||c|c||c|c|}\hline Method & \multicolumn{4}{c||}{\emph{\footnotesize{covered}}} 
 & \multicolumn{4}{c|}{\emph{\footnotesize{perforated}}}
 \\ \hline
  \footnotesize{Reference-based MI} & \scriptsize{Mean} & \scriptsize{SE} 
  & \scriptsize{Mean} & \scriptsize{SE}
  & \scriptsize{Mean} & \scriptsize{SE}
  & \scriptsize{Mean} & \scriptsize{SE}
 \\ \hline
 J2R only & 2.18 & 1.13 &      &       & 2.17 & 1.13 &    &     \\
 J2R + observed Off-treatment data& 2.28 & 1.05 &      &       & 2.39 & 1.05  &    &     \\ \hline
Compliance-based MI & \multicolumn{2}{c|}{\scriptsize{\emph{historic}}}
 & \multicolumn{2}{c||}{\scriptsize{\emph{current}}}
 & \multicolumn{2}{c|}{\scriptsize{\emph{historic}}}
 & \multicolumn{2}{c|}{\scriptsize{\emph{current}}}
 \\ \hline
 \%MISTEP macro  & 2.32 &  1.10 & 2.34 &  1.07 & 2.45\footnotemark[1] & 
 1.08\footnotemark[1] & 2.74\footnotemark[1]  & 1.05\footnotemark[1] 
 \\
 Proc MI & 2.32 &  1.10 & 2.34 &  1.07 & 2.44\footnotemark[2] & 
 1.10\footnotemark[2] & 2.39\footnotemark[2] & 1.06\footnotemark[2]  
 \\
 Proc BGLIMM  & 2.31 &  1.10 & 2.35 &  1.06 &   2.75\footnotemark[3]     & 
 2.39\footnotemark[3] & 2.52\footnotemark[3] & 1.05\footnotemark[3]
 \\
 \hline
 \multicolumn{9}{|l|}{\scriptsize{J2R with \emph{historic} and \emph{current} extensions with varying
 prior variance} } \\ \hline
 Var=1   & 2.28 &  1.05 & 2.29 &  1.05 & 2.38     &  1.04     & 2.42 & 1.04   \\
 Var=10  & 2.31 &  1.06 & 2.33 &  1.06 & 2.40     &  1.08     & 2.49 & 1.05   \\
 Var=40  & 2.32 &  1.08 & 2.35 &  1.06 & 2.41     &  1.16     & 2.51 & 1.05   \\
 Var=160 & 2.32 &  1.09 & 2.36 &  1.06 & 2.44     &  1.45     & 2.52 & 1.05   \\
 Var=1000& 2.32 &  1.10 & 2.36 &  1.06 & 2.63     &  2.80     & 2.52 & 1.05   \\
\hline
 \multicolumn{9}{|l|}{\scriptsize{CIR with \emph{historic} extension with varying
 prior variance} } \\ \hline
 Var=1   & 2.41 &  1.04 &  &   & 2.42     &  1.04     &  &    \\
 Var=10  & 2.39 &  1.06 &  &   & 2.41     &  1.08     &  &    \\
 Var=40  & 2.36 &  1.08 &  &   & 2.41     &  1.16     &  &    \\
 Var=160 & 2.34 &  1.09 &  &   & 2.44     &  1.45     &  &    \\
 Var=1000& 2.32 &  1.10 &  &   & 2.63     &  2.79     &  &    \\
\hline 
\multicolumn{9}{|l|}{\scriptsize{MAR+2 with \emph{historic} extension  with varying
 prior variance} } \\ \hline
 Var=1   & 2.53 &  1.06 &  &   & 2.43  &  1.07  &  &    \\
 Var=10  & 2.44 &  1.07 &  &   & 2.31  &  1.09   &  &    \\
 Var=40  & 2.38 &  1.08 &  &   & 2.24  &  1.17   &  &    \\
 Var=160 & 2.34 &  1.09 &  &   & 3.24  &  1.48   &  &    \\
 Var=1000& 2.33 &  1.10 &  &   & 2.43  &  2.81   &  &    \\
 \hline
 \end{tabular}
\\ \footnotesize{
1: Those patient visits which cannot be imputed are removed.\\
2: Proc MI sets un-estimable parameters to zero before imputing.\\
3: Prior N(0,1000) for all fixed effects as some parameters non-estimable.}
\end{table}

\section{Discussion}
\label{sec:Discussion}

In this article we have considered the analysis of clinical trials with quantitative longitudinal outcome data, where not all patients remain on treatment and complete follow-up to the end of the study and a treatment policy strategy is of interest for addressing treatment-withdrawal. Compliance-based multiple imputation using off-treatment data (also referred to as retrieved drop out multiple imputation) has been recommended for estimation in this setting. However, a perforated data structure may render such imputation methods infeasible since imputation model parameters may be inestimable. With several follow-up time points a perforated data structure becomes increasingly likely.   We have proposed a novel imputation model parameterisation for off-treatment multiple imputation which can be implemented with limited observed off-treatment data.

In previous work others have proposed the use of retrieved drop out multiple imputation methods as preferable in this setting and where not possible, due to convergence issues, reverting to simpler multiple imputation methods such as
the return to baseline (RTB) method or the washout method  \cite{Wang23}. Our method improves upon the latter fall-back approaches by utilising partially observed off-treatment data combined with a reference-based model. Appealingly a single non-adaptive analysis plan can be presented when rates of retrieved off-treatment data are unknown. This is preferred to an adaptive analysis approach dependent on the realised data structure, which may be sub-optimal.

It has previously been shown that reference-based multiple imputation provides approximately information anchored inference \cite{Cro19}. That is, Rubin’s variance estimator for the treatment effect ensures that the loss of information due to missing data  is approximately the same as analysis under the MAR assumption for a broad range of commonly used reference-based alternatives. It has also been shown that delta-based multiple imputation with a fixed delta is also information anchored; when a prior is incorporated on delta the variance of the MI treatment effect, estimated using Rubin's rules, will incorporate the additional variance on delta (because the variation in delta increases the between-imputation variance)\cite{Cro19}. Since the proposed method consists of a reference based model plus some additional model parameters with a prior mean of zero and specified variance, it can be inferred that the variance estimator for the treatment effect will be approximately information anchored, with  an additional loss of information that is dependent on the prior variance assumed for the extended model parameter. Our results in table \ref{Results} show how increasing the prior variance for the extended model increases the variance for the estimated treatment effect by a small amount.   
If the deviation process is not at random, then it is still possible by using
off-treatment data to make an unbiased estimate of the treatment policy
estimand. However if the missingness mechanism is not missing at random then 
other approaches are needed to generate the missed data. The main driver for such MNAR 
data could be that those who immediately leave the study are in some way different from those remaining.
By conditioning on proxies for these factors the MAR approximation should improve. The MAR assumption at trial withdrawal can be made more likely by including covariates that tell us about treatment deviation (e.g. Reason for withdrawal) or by stratifying the compliance model by treatment.  Or a delta-based multiple imputation approach may be useful. Here delta is applied following trial withdrawal rather than in the core model which handles post-deviation irregularity.

We have used an open access part computer generated data set to demonstrate how easy it is to implement the proposed method and enable readers to replicate. As such it is important not to draw conclusions from the data about the types of pattern that can be expected in terms of treatment withdrawal and/or study withdrawal. It is important that therapeutic areas make publicly available summary data on the ability to collect such off-treatment data and the likely trajectories for those who do not comply.

In these example data sets all those patients who come off treatment either immediately stop further observation or are observed right to the end of the study. There are no patients with some but only some of their follow-on data available. Patients who have some but not all post-deviation data will have imputation condition on both their on-treatment and their off-treatment data This will generate more complex scenarios than those faced in our two examples. For the \emph{current} approach the fact that the $\alpha_{12}$ parameter is not estimable is unimportant in our examples as it is not actually used for imputation. Patients who need imputation at the final visit have no off-treatment data and so knowing the mean at visit 2 is not needed. However if there were patients who deviate and have only some off-treatment data then knowing the value for $\alpha_{12}$ would be potentially important. The basic idea of providing a mildly informative prior for the deviations from a core RBI model should be even more useful in the more realistic setting of partial off-treatment data within patient.

We have assumed there are no interim missing data although we can reasonably expect to experience such issues in practice. Both on-treatment and off-treatment intermediate missing data could be handled with MAR Multiple Imputation. When data are missing for other reasons beyond the intercurrent events handled by use of treatment policy estimand, such as death, other strategies may be needed. Multiple imputation under other scenarios may allow a joined-up approach to the disparate types of intercurrent event.

We have assumed independence of the priors for the deviations. If there are several visits within a study, we may expect the deviations to be in the same direction at adjacent visits. So one might want to be build in some positive correlation between adjacent deviations within treatment. This is relatively easy to do within the code we propose, simply adding extra correlation rows to the priors data set. A correlation as large as 0.5 might be suitable, but we have not explored this yet.

Rather than using BGLIMM, the \emph{historic} approach can be implemented in a stepwise way visit by visit using standard Bayesian univariate regression. Such software is generally available and can also be programmed oneself as the required priors are all conjugate leading to direct sampling rather than involving complex MCMC algorithms. One advantage of the marginal BGLIMM approach is that it can automatically handle intermediate missing values under MAR. An advantage of a sequential implementation is that the RBI idea of switching covariance matrix at deviation \cite{carpenter13} could be implemented quite simply by stratifying at each visit the regression on previous observed values by whether the observed value was on or off treatment. For the \emph{current} approach the Bayesian univariate regressions need to be based on previous residuals. Other than BGLIMM, the only SAS Bayesian regression procedure which allows user-specified priors is GENMOD. However the BY statement in this procedure currently uses the same seed for each BY group, so messy macro looping is required. Also the imputed values are not directly available as in BGLIMM.

Overall we have illustrated how Multiple imputation provides a computationally practical method for inference in this framework. We propose a core reference-based model is combined with a compliance model, using both on- and off- treatment data to form an extended model for imputation. SAS code has been provided to readily enable implementation of the proposed method. The EIWG is conducting research to compare different modeling approaches
to estimate estimands incorporating treatment policy strategies for intercurrent events in continuous endpoints and  plans to provide recommendations on appropriate approaches based on the simulation results\cite{fletcher22}. \\

\emph{Data availabilty statement}

The data analysed in this article are open access and available at https://www.lshtm.ac.uk/research/centres-projects-groups/missing-data\#dia-missing-data\cite{SWGEM}.

\emph{Funding statement}

Suzie Cro is funded by an NIHR advanced research fellowship (NIHR300593). James Carpenter is funded by the Medical Research Council MC UU 00004/07.

\emph{Conflict of interest disclosure}

James Roger is a previous employee of GSK, J\&J. Livedata Process and Livedata (UK) Ltd; he is a shareholder in GSK and his wife is a shareholder in AZ and GSK. 

\emph{Ethics approval discolsure}

Not applicable.

\emph{Patient consent statement}

Not applicable.

\emph{Permission to reproduce material from other sources}

Not applicable.

\emph{Clinical trial registration}

Not applicable.

 \bibliography{Manuscript}

\section{Appendix}

This appendix shows how to implement this approach in SAS and generate an Imputation data set, ImputedData.

\begin{itemize}
\item Generate a vertical data set InputData with the required variables.\\
It is assumed that the data set is in vertical format (One patient-visit per record) already indexed by patient identification, Patient, and visit number Visit. Also that there are two linked variables, variable OffT taking values 0 or 1, where 1 indicates the patient has come off treatment and LastVis which is the visit index of the patient's last on-treatment visit. The LastVis value relates to the patient and has the same value for all records within a patient, whilst the OffT is a numeric indicator variable which varies from visit to visit following the definition.
\begin{itemize}
\item[] if Visit > LastVis then OffT=1; else OffT=0;
\end{itemize}
The variable Pattern is a derived factor also defined at the visit level taking the current visit index whilst still on treatment and remaining at the index value of the last visit on-treatment for that and all remaining visits.
\begin{itemize}
\item[] if Visit > Lastvis then Pattern=Lastvis; else Pattern=Visit;
\end{itemize}
Also the variable Trt holds the treatment levels, 0 and 1 here, and J2R is a factor holding the treatment level whilst on treatment and the reference level 0 whilst off treatment.
\begin{itemize}
\item[] if Trt= 0 then J2R=0; else if visit<=Lastvis then J2R=1; else J2R=0;
\end{itemize}
\item Here we fit the Bayesian model with Change as the outcome variable and baseline covariate Basval using J2R+\emph{historic} as method. 1000 samples are created with a thinning of 5. Include the trace statement to see diagnostic plots for the MCMC process.
\begin{itemize}
\item[]proc bglimm data=InputData seed=21 outpost=MyPosterior nmc=5000 thin=5 ; * plots=trace;\\
class Patient Trt Visit J2R Pattern;\\
model Change=  OffT*Trt*Visit*Pattern  J2R*Visit  Basval*Visit  / noint CPRIOR=NORMAL(Input=MyPriors);\\
repeated Visit / subject=Patient type=un;\\
run;
\end{itemize}
If a "delta" approach is required for the core model then this can implemented using the OFFSET option on the model statement, along with a variable holding the required delta values. But beware that the BGLIMM procedure may need a Hot Fix for this to give correct values. A workaround for this when using the identity link is to subtract the offset from the response variable before running BGLIMM and then add it back to the imputed values.

\item The priors for the parameters in the term OffT*Trt*Visit*Pattern need to have been set in the data set MyPriors, and running the BGLIMM procedure as above without the CPRIOR= option will provide their names. Then code like this can be used to set the zero mean and variance at 40 thus.
\begin{itemize}
\item[]options validvarname=any;\\
Data Mypriors;\\
length \_Type\_ \$4 ;\\
\%macro dummy();\\
\%do i=0 \%to 1; \%do j=1 \%to 4; \%do k=1 \%to \&j;\\
            "OffT*Trt \&i.*visit \&j.*Pattern \&k"n = 0;\\
\%end; \%end; \%end;\\
\_Type\_="Mean"; output;\\
\%do i=0 \%to 1; \%do j=1 \%to 4; \%do k=1 \%to \&j;
            "OffT*Trt \&i.*visit \&j.*Pattern \&k"n = 40;\\
\%end; \%end; \%end;\\
\%mend dummy; \%dummy;\\
\_Type\_="Var "; output\\
run;
\end{itemize}
The validvarname option is required so that the variable names can include the spaces required by BGLIMM. The dummy macro is required to generate the names as the SAS language does not allow macro do loops within open code.

\item Then the 1000 sampled posterior values for the missing data are merged back with the original data to create 1000 imputed data sets. This can be done using the \%BGI macro from the DIA web site or if required use code generated by the \%BGI macro as template code to directly carry out a transpose of the MyPosterior data set and then merge it back into th original data set. The required call to the macro is
\begin{itemize}
\item[]\%BGI(Data=InputData, Post=MyPosterior, Out=ImputedData, Response=Change );
\end{itemize}
\item For other core models it is necessary to set up a treatment by visit array TrtbyVis in the InputData data set to specify the required RBI method. First declare the array and zero the elements.
\begin{itemize}
\item[]array TrtbyVis[2,4] ;\\
do t=1 to 2; do v=1 to 4;\\
TrtbyVis[t,v]=0;\\
end; end;
\end{itemize}
then for CIR use,\ldots
\begin{itemize}
\item[]if Trt= 0 then TrtbyVis[Trt+1,Visit]=1; \\
else if Visit<=Lastvis then TrtbyVis[Trt+1,Visit]=1;\\
else do;\\
TrtbyVis[1,Visit]=1;\\
TrtbyVis[Trt+1,Lastvis]=1;\\
TrtbyVis[1,Lastvis]=-1;\\
end;
\end{itemize}
or for LMCF use \ldots
\begin{itemize}
\item[]if Visit<=Lastvis then TrtbyVis[Trt+1,Visit]=1;\\
else TrtbyVis[Trt+1,Lastvis]=1
\end{itemize}
and one could do J2R this way thus \ldots
\begin{itemize}
\item[]if Trt= 0 then TrtbyVis[Trt+1,Visit]=1; \\
else if Visit<=Lastvis then TrtbyVis[Trt+1,Visit]=1;\\
else TrtbyVis[1,Visit]=1;\\
end;
\end{itemize}
and finally replace J2R by TrtbyVis1-8 in the model statement
\begin{itemize}
\item[]model Change=  Offt*Trt*visit*Pattern TrtbyVis1-TrtbyVis8 Basval*Visit  / noint 
\end{itemize}
\end{itemize}

\end{document}